\documentclass[proc]{edpsmath}
\usepackage{amsmath,amssymb}
\usepackage{color}
\newtheorem{remark}{Remark}
\include{graphics}
\usepackage{mathrsfs}
\usepackage{psfrag}
\usepackage{epsfig}
\graphicspath{{pix/}}

\begin{document}

%%-----------------------------
%%      the top matter
%%-----------------------------
\title{Statistical methods for critical scenarios in aeronautics}
%\thanks{Safety Line}\thanks{Labex AMIES}\thanks{CNRS Libanais}% At most 5 thanks
%
\author{Houssam Alrachid}\address{Universit\'e Paris Est, CERMICS, Ecole des Ponts and INRIA, 6 \& 8 avenue Blaise Pascal, 77455 Marne-la-Vall\'ee Cedex 2, France; \email{alrachih@cermics.enpc.fr \&\ ehrlachv@cermics.enpc.fr \&\ lelievre@cermics.enpc.fr} }
\secondaddress{Universit\'{e} Libanaise, Ecole Doctorale des Sciences et de Technologie (EDST), Campus Universitaire de Rafic Hariri, Hadath, Lebanon.}
\author{Virginie Ehrlacher}\sameaddress{1}
\author{Alexis Marceau}\address{Safety Line, Hall B 15 rue Jean Baptiste Berlier 75013 Paris, France; \email{contact@safety-line.fr}}
\author{Karim Tekkal}\sameaddress{3}

%
%\dedicated{\it Dedicated to Maurice Dupont} %if necessary
%
\begin{abstract} 
We present numerical results obtained on the CEMRACS project Predictive SMS proposed by Safety Line. The goal of this work was to elaborate a purely statistical method in order to reconstruct the deceleration profile 
of a plane during landing under normal operating conditions, from a database containing around $1500$ recordings. The aim of Safety Line is to use this model to detect malfunctions of the braking system of the plane from deviations of the measured deceleration profile of the plane to the one predicted by the model. This yields to a multivariate nonparametric regression problem, which we chose to tackle using a Bayesian approach based on the use of gaussian processes similar to the one presented in \cite{ed}. We also compare this approach with other statistical methods.  
 \end{abstract}
\begin{resume} 

Nous pr\'esentons des r\'esultats num\'eriques obtenus sur le projet CEMRACS Predictive SMS propos\'e par Safety Line. L'objectif de ce travail \'etait d'\'elaborer une m\'ethode purement statistique afin de reconstruire le profil de d\'ec\'el\'eration d'un avion durant son atterissage, \`a partir d'une base de donn\'ees contenant \`a peu pr\`es $1500$ enregistrements. Le but de Safety Line est d'utiliser ce mod\`ele pour d\'etecter des anomalies du syst\`eme de freinage de l'avion \`a partir de l'\'ecart entre le profil de d\'ec\'el\'eration de l'avion mesur\'e et celui pr\'edit par le mod\`ele. Ceci m\`ene \`a un probl\`eme de r\'egression multivari\'e non param\'etrique que nous avons choisi de traiter via une approche bay\'esienne utilisant des processus gaussiens similaire \`a celle pr\'esent\'ee dans  \cite{ed}. Nous comparons \'egalement cette approche avec d'autres m\'ethodes statistiques classiques.
 \end{resume}
\maketitle
%%-----------------------------
%%      your text
%%-----------------------------
\section*{Introduction}

Safety Line is a company that offers innovative solutions (software and statistical analysis) for risk management in the field of air transport (airlines, maintenance organizations, airports ...). The main expertise of Safety Line relies in hazard identification and risk assessment, assurance and safety promotion.\\

The objective of Safety Line for this CEMRACS project is to improve its technical modeling of plane systems. To monitor the proper functioning of a 
given system, their overall approach is to follow the time evolution of an indicator of the state of the system and detect deviations from the expected 
behaviour, which could be indicative of a malfunction. For this, the main challenge is to estimate as precisely as possible and at any time 
the value of this indicator in normal operating conditions. In this project, we are interested in evaluating the state of the plane braking system. 
To this aim, we chose to focus on the indicator given by the deceleration force of the plane during landing, from the moment when the plane wheels 
touch the ground up to the moment when the plane leaves the track. Indeed, the deceleration force of the plane is the sum of an aerodynamic component and 
a component related to the brakes. There are no wear on the aerodynamic component, unlike that of the brakes which varies according to the state of the system.\\
To this aim, we have at our disposal a database containing the measurements recorded over $1505$ landings. For each landing, different time-dependent quantities 
are available such as the deceleration profile, the angle of the brake level manipulated by the pilot, the speed of the plane...\\
The goal of this project is to build a purely statistical model from the available data, which can be used to reconstruct the deceleration force profile of 
the plane given a new set of input quantities. To achieve this task, we chose a Bayesian approach based on gaussian processes, inspired from ideas of~\cite{ed} 
and present the results we obtained with this approach. We also compared our strategy with other regression models which were already used by Safety Line, 
such as linear regression methods or random forests. The approach using gaussian processes seems to perform significantly better than all these other methods.\\

In Section~\ref{sec 1}, we detail the structure of the database we trained our model on, before moving to the presentation of the Bayesian approach with Gaussian processes in Section~\ref{sec 2}. The numerical results we obtained are commented in Section~\ref{sec 3}. In Section~\ref{disc}, we provide some discussions about other methods that take into account time series influence in the prediction model.\\

\section{Data provided by Safety Line}\label{sec 1}

    \subsection{Problem presentation}\label{sec 1.1}

As announced in the introduction, we are interested in reconstructing the time-dependent profile in the  deceleration force of the plane during landing, 
which is a good indicator of the state of the braking system, from given input quantities.\\
The objective of building this statistical model is to detect malfunctions of the plane from the deviation of the recorded deceleration force 
profile to the predicted one. In order to do so, ideally, our statistical model should be trained on data recorded for planes whose braking systems are new, 
or at least in a good state. Unfortunately this is a piece of information we did not have access to. Thus, we trained our model on the data we had at hand, 
the main objective being to demonstrate the feasibility and potential of our approach.\\
However, the state of the braking system of most planes recorded being hopefully good for most planes in average, it is reasonable to think that this data 
will nevertheless be sufficient to provide information on the mean behaviour of deceleration profiles of planes during landing, and that it should 
give useful indications in order to detect malfunctions of their braking system.\\
At this point, a first difficulty was to clearly define the set of inputs our statistical model will be built on. Indeed, for each landing, a huge amount 
of data is recorded and we have to select the quantities that are meaningful for the reconstruction of the deceleration force profile. We chose to select input quantities 
that enable us to evaluate the different forces which act on the plane during its landing.\\

All time-dependent data are provided discretely with one measurement per second during a period of $T=100$ seconds. Thus, for any quantity $a$, its profile during 
landing will be characterized by a vector of size $T+1$, $(a^t)_{0\leq t\leq T}\in \mathbb{R}^{T+1}$, where $a^t$ denotes the measurement of the quantity $a$ at 
time $t\in\{0,1,2,...,T\}$.
  
    \subsection{Input quantities}\label{sec 1.2}
    
We consider the following input data:\\
  
  \begin{itemize}
 \item [1$\cdot$] the weight of the plane $m$;\\
 \item [2$\cdot$] the initial kinetic energy of the plane $e= \frac{1}{2}m(v^0)^2 $, where $ v^0 $ is the initial speed of the plane; \\
 \item [3$\cdot$] the speed of the plane during the landing $v=(v^t)_{0\leq t\leq T}$;\\
 \item [4$\cdot$] the thrust force $ p=(p^t)_{0\leq t\leq T}$,  where for all time $t\in \{0,...,T\}$, $p^t$ is evaluated as the product of the square of 
the speed of the plane $(v^t)^2$ times the level of the reverse throttle;\\
 \item [5$\cdot$] the vector $b= (b^t)_{0\leq t\leq T}$; at any time $t\in \{0,...,T\}$, $b^t$ is equal to the product of the speed times the angle of the 
brake level, thus $b$ gives an indication of the braking force during landing; 
 \item [6$\cdot$] the drag force $\delta=(\delta^t)_{0\leq t\leq T}$, which is a function of $(v^t)^2$ at any time $t\in\{0,...,T\}$.
\end{itemize}
All these quantities, except the weight and the initial kinetic energy, are time-dependent functions.
  \begin{remark}
  Note that rebuilding the deceleration from only  the speed of the plane is not that obvious. In fact, the sensors are not optimal (because of the noise),
 so that the derivative of the  speed signal is not equal to the measured deceleration profile. This leads to additional difficulties.
  \end{remark}
  
  \subsection{Output quantity}\label{sec 1.3}

The output quantity we wish to reconstruct is the deceleration force of the plane $\gamma$, given as a vector 
$(\gamma^t)_{0\leq t\leq T}$, where for all $0\leq t\leq T$, $\gamma^t$ is equal to the product of the deceleration times the mass of the plane.\\
  
  \subsection{Database}  \label{sec 1.4}
  We consider a training data set which contains the recordings related to $n_{ob}=1505$ different landings. In all the rest of the document, 
the superscript $i$ $(1\leq i\leq n_{ob})$ refers to the label of the landing recorded in the database; the index $k$ refers to the type of input quantity 
among the $6$ considered and presented in Section~\ref{sec 1.2} (weight, initial kinetic energy, speed, thrust force, braking force and drag force); lastly, the superscript 
$t$ refers to the instant of the measurement ($0\leq t\leq T$).\\

More precisely, the data recorded for the $i^{th}$ landing at a time $t\in\{0, ..., T\}$ consists of:\\
\begin{align*}
x^{i,t}_{d,1} & =  m^i, \\
x^{i,t}_{d,2} & =  e^i,\\
x^{i,t}_{d,3} & =  v^{i,t},\\
x^{i,t}_{d,4} & =  p^{i,t},\\
x^{i,t}_{d,5} & =  b^{i,t},\\
x^{i,t}_{d,6} & =  \delta^{i,t},\\
y^{i,t}_{d}    & = \gamma^{i,t},
\end{align*}
where we use the same quantities as those introduced in Sections~\ref{sec 1.2} and~\ref{sec 1.3}.\\ 
We also denote by $y^i_d=(y^{i,t}_d)_{0\leq t\leq T}$ and $ x^i_d=(x^{i,t}_{d,k})_{0\leq t\leq T,\;1\leq k\leq 6}$. Thus, for all $1\leq i\leq n_{ob}$, 
$y^i_d\in \mathbb{R}^{T+1}$ and $x^i_d\in \mathbb{R}^{6(T+1)}$. The training data set can then be written as $D = \left\{ (y_d^i, x_d^i), i = 1, \cdots, n_{ob}\right\}$. 
We also denote by $X_d = (x_d^i)_{1\leq i \leq n_{ob}}\in \mathbb{R}^{n_{ob}\times(6(T+1))}$ 
(respectively $Y_d = (y_d^i)_{1\leq i \leq n_{ob}}\in \mathbb{R}^{n_{ob}\times(T+1)}$) the set of input (respectively output) quantities of the database. 
For all $0\leq t\leq T$, we also define $Y_d^t= (y_d^{i,t})_{1\leq i \leq n_{ob}}\in \mathbb{R}^{n_{ob}}$.

  \section{Construction of the regression model}\label{sec 2}
    
  \subsection{The Bayesian approach and gaussian processes}\label{sec 2.1}

We use here a Bayesian approach presented in~\cite{ed}.\\
Let $(\Omega,\mathcal{A},\mathbb{P})$ be a probability space. All input and output quantities are considered as random vectors.\\
Let $t\in \{0,...,T\}$ be a fixed time. We assume that, for all $\textbf{x}\in \mathbb{R}^{6(T+1)}$ (random) value of the input quantities, 
the associated observed value of the output quantity at time $t$, denoted by $\textbf{y}^t\in \mathbb{R}$, can be written as:\\
\begin{equation}\label{gp1}
\textbf{y}^t=\mathcal{F}^t(\textbf{x})+\textbf{g}^t,
\end{equation}
where $\mathcal{F}^t$ is a (random) function $\mathcal{F}^t:\mathbb{R}^{6(T+1)} \rightarrow \mathbb{R}$ such that $\mathcal{F}^t(\textbf{x})$ 
is equal to the true value of the output quantity at time $t$ for the input $\textbf{x}$, and where $\textbf{g}^t\in \mathbb{R}$ is a random variable modeling the error due to the noise made on the measurement of the output quantity a time $t$. More generally, for a set of $n_{ob}$ random inputs, $\textbf{X}=(\textbf{x}^i)_{1\leq i \leq n_{ob}}$, if we denote by $\textbf{Y}^t=(\textbf{y}^{i,t})_{1\leq i \leq n_{ob}}\in \mathbb{R}^{n_{ob}\times 1}$ the set of measured associated output values, $\mathcal{\textbf{F}}^t=(\mathcal{F}^t(\textbf{x}^i))_{1\leq i \leq n_{ob}}\in \mathbb{R}^{n_{ob}\times 1}$ and $\textbf{G}^t=(\textbf{\textbf{g}}^{i,t})_{1\leq i \leq n_{ob}}$ the set of noises made on the measurement of the output quantities, we have\\
\begin{equation}\label{gp2}
\textbf{Y}^t=\textbf{F}^t+\textbf{G}^t.
\end{equation}

In a Bayesian framework, the law of $\textbf{Y}^t|\textbf{X},\textbf{F}^t$ is called the \textit{likelihood} and is chosen a priori. 
It is directly related to the choice of the law of the random vector $\textbf{G}^t$ modeling the noise made on the measurements of the output quantity. In the sequel, 
we will assume that the variables  $(\textbf{g}^{i,t})_{1\leq i \leq n_{ob}}$ are independent, identically distributed, centered gaussian variables with 
variance $\sigma^2$, so that $\textbf{G}^t$ is a random gaussian vector of law
$$\textbf{G}^t\sim \mathcal{N}(0,\sigma^2I_{n_{ob}}), $$
where $I_{n_{ob}}$ denotes the identity matrix of $\mathbb{R}^{n_{ob}\times n_{ob} }$. Thus,
$$\textbf{Y}^t|\textbf{X},\textbf{F}^t \sim \mathcal{N}(\textbf{F}^t,\sigma^2I_{n_{ob}}).$$
Besides, the law of the function value $\textbf{f}^t=\mathcal{F}^t(\textbf{x})$ conditioned to the knowledge of the value of the input $\textbf{x}$ 
is called the \textit{prior distribution} and is also chosen a priori. We use here a model where $\textbf{f}^t|\textbf{x}$ is assumed to be a gaussian process 
characterized by its mean $\mu:\mathbb{R}^{6(T+1)} \rightarrow \mathbb{R}$ and covariance function 
$\kappa:\mathbb{R}^{6(T+1)} \times \mathbb{R}^{6(T+1)}\rightarrow \mathbb{R}$, i.e.
$$\textbf{f}^t|\textbf{x} \sim\mathcal{GP}(\mu(\textbf{x}), \kappa(\textbf{x}, \textbf{x'})).$$
In particular, this implies that $\textbf{F}^t|\textbf{X}$ is a random gaussian vector of size
 $n_{ob}\times 1$, of mean $M_\textbf{X}\in \mathbb{R}^{n_{ob}\times 1 }$ and covariance matrix  
$K_{\textbf{X},\textbf{X}}\in \mathbb{R}^{n_{ob}\times n_{ob} }$ where 

\begin{align}
M_\textbf{X} & :=(\mu(\textbf{x}^i) )_{1\leq i \leq n_{ob} \,\mbox{and} \label{mu}}\\
 K_{\textbf{X},\textbf{X}} & =(\kappa(\textbf{x}^i,\textbf{x}^j) )_{1\leq i,j \leq n_{ob}}.
\end{align}
Thus, $\textbf{F}^t|\textbf{X} \sim \mathcal{N}(M_{\textbf{X}}, K_{\textbf{X}, \textbf{X'}})$.
In the sequel, we assume that $\mu=0$ and that the covariance function (or kernel) $\kappa$ is chosen as a squared exponential covariance function defined by:\\
for all $x^r =(x^{r,t}_{k})_{1\leq k\leq 6,\;0\leq t\leq T}\in\mathbb{R}^{6(T+1)}$, $x^s =(x^{s,t}_{k})_{1\leq k\leq 6,\;0\leq t\leq T}\in\mathbb{R}^{6(T+1)}$,\\
\begin{equation}\label{ker}
\kappa(x^r,x^s):=\tau^2 {\rm exp}\left [  \displaystyle \sum_{k=1}^6\frac{1}{2l_k} \|x^{r,.}_{k}-x^{s,.}_{k}\|^2_{\mathbb{R}^{(T+1)}} \right]
\end{equation}
where $\|\cdot\|_{\mathbb{R}^{(T+1)}}$ denotes the Frobenius norm on vectors of dimension $T+1$.\\
The parameters $\sigma,$ $\tau$, $l_k\,(1\leq k\leq 6)$, which the prior distribution and likelihood depend on, are positive real numbers called \itshape hyperparameters\normalfont,
 and their values depend a priori on the time $t\in\{0,...,T\}$. They have to be chosen in an appropriate way which is detailed in Section~\ref{sec 2.3}.

 \begin{remark}
At this point, we would like to comment on the simplistic choice we made assuming that the mean function $\mu$ should be zero, and in the particular form 
of the kernel function $\kappa$ we introduced above. 

\medskip

Of course, this choice is not the only possibility and one could think for instance to borrow ideas from kriging methods in order to obtain 
a better guess of this mean function. The universal kriging technique is an example of such a method which could be used
to evaluate the function $\mu$. Indeed, in this case, the function $\mu$ is approximated by
\begin{equation}\label{mukri}
\mu(\textbf{x})=\sum_{l=1}^L \beta_l h_l(\textbf{x}),
\end{equation}
for some $L\in\mathbb{N}^*$, where $h_1,...,h_L:\mathbb{R}^{6(T+1)} \to \mathbb{R}$ is a set of a priori fixed basis functions, and $(\beta_1,...,\beta_L)\in \mathbb{R}^L$ are
real coefficients which can be viewed as an additional set of hyperparameters and which are to be determined from the database we have at hand. 
However, in our case, because of the high-dimensional character of the input quantities of our database ($6(T+1) = 606$ variates), 
the choice of a meaningful 
set of basis functions and the identification of associated parameters 
is a quite intricate task. Indeed, even if we chose a simplistic linear regression model, we would have to fit $606$ additional hyperparameters. 
It would be interesting though to test if the choice of a better mean function $\mu$ could help in improving the results we obtained. 
As announced above, the numerical results presented below were obtained in the simple case where the mean function $\mu$ is assumed to be $0$.
\end{remark}

\subsection{Reconstruction of the value of the output quantity from a new set of input values }\label{sec 2.2}

Assume for now that the values of the hyperparameters $\sigma,$ $\tau$, $l_k$ ($1\leq k\leq 6$), have been chosen for the time $t$.\\
Let $n_{test}\in \mathbb{N}^*$ and $X_*:=(x_*^i)_{1\leq i \leq n_{test}} \in \mathbb{R}^{n_{test}\times (6(T+1))}$ be a set of $n_{test}$ new input vectors such that a priori $X_*$ is not included 
in the set $X_d$ of input values of the database. We present in this section 
how the set of the values of the deceleration of the plane at the time $t$ for each input vector, $F_*^t = (f_*^{i,t})_{1\leq i \leq n_{test}}\in \mathbb{R}^{n_{test}}$, 
can be reconstructed using the regression model based on gaussian processes.\\

Let us consider a test random vector of input quantities $\textbf{X}_* = \left(\textbf{x}^i_*\right)_{1\leq i \leq n_{test}}\in\mathbb{R}^{n_{test} \times (6(T+1))}$ 
and denote by $\textbf{F}_*^t = \left(\textbf{f}_*^{i,t}\right)_{1\leq i \leq n_{test}} \in \mathbb{R}^{n_{test}\times 1}$ where for all $1\leq i \leq n_{test}$, 
$\textbf{f}_*^{i,t}=\mathcal{F}^t(\textbf{x}^i_*)$ is the "true" output value for the input vector $\textbf{x}_*^i\in \mathbb{R}^{6(T+1)}$.\\
The joint distribution of the previously observed target values $\textbf{Y}^t$ and the randon vector $\textbf{F}_*^t$
 can be written as (using the gaussian process model introduced in the preceding section ): 

 $$\left[
 \begin{array}{c}
\textbf{Y}^t\\
\textbf{F}_{*}^t\\
\end{array}\right]
\sim  \mathcal{N} \left(
0,
\left[
 \begin{array}{cc}
K_{\textbf{X},\textbf{X}}+ \sigma^2I_{n_{ob}} & K_{\textbf{X},\textbf{X}_*}\\
K_{\textbf{X}_*,\textbf{X}} & K_{\textbf{X}_*,\textbf{X}_*}\\
\end{array}\right]
\right),
$$
where 
$$K_{\textbf{X},\textbf{X}_*}=(\kappa(\textbf{x}^i,\textbf{x}^j_*))_{1\leq i \leq n_{ob}, \; 1\leq j \leq n_{test}}\in\mathbb{R}^{n_{ob}\times n_{test}},$$
$$K_{\textbf{X}_*,\textbf{X}}=(\kappa(\textbf{x}^i_*,\textbf{x}^j))_{1\leq i \leq n_{test}, \; 1\leq j \leq n_{ob}}\in\mathbb{R}^{n_{test}\times n_{ob}},$$
$$\mbox{and} \, K_{\textbf{X}_*,\textbf{X}_*}=(\kappa(\textbf{x}^i_*,\textbf{x}^j_*))\in\mathbb{R}^{n_{test}\times n_{test}}.$$
We thus obtain the law of $\textbf{F}_*^t|\textbf{X},\textbf{Y}^t,\textbf{X}_*$ which reads 
$$\textbf{F}_*^t|\textbf{X},\textbf{Y}^t,\textbf{X}_*\sim\mathcal{N}(\overline{\textbf{F}}_*^t, \textbf{S}_*^t) $$

where
\begin{align*}
\overline{\textbf{F}}_*^t & =  \mathbb{E}\big[ \textbf{F}_*^t|\textbf{X},\textbf{Y}^t,\textbf{X}_* \big ], \\
& =  K_{\textbf{X}_*,\textbf{X}}\big ( K_{\textbf{X},\textbf{X}}+ \sigma^2I_{n_{ob}} \big )^{-1}\textbf{Y}^t,\\
\textbf{S}_*^t    & = K_{\textbf{X}_*,\textbf{X}_*}-K_{\textbf{X}_*,\textbf{X}} \big [   K_{\textbf{X},\textbf{X}}+\sigma^2I_{n_{ob}}\big ] ^{-1} K_{\textbf{X},\textbf{X}_*} 
\end{align*}

For all $1\leq i \leq n_{test}$, the vector of the reconstructed values of the deceleration of the plane at time $t$, $F_*^t\in \mathbb{R}^{n_{test}\times 1}$ 
associated to the set of input quantities 
$X_*\in\mathbb{R}^{6(T+1)}$ is then given by 
\begin{align*}
F_*^{t} & =  \mathbb{E}\big[ \textbf{F}_*^{t}|\textbf{X}=X_d,\textbf{Y}^t=Y_d,\textbf{X}_*=X_* \big ]=  K_{X_*,X_d}\big [ K_{X_d,X_d}+ \sigma^2I_{n_{ob}} \big ]^{-1}Y_d^t,\\
\end{align*}
Thus, for all $1\leq j \leq n_{test}$, $f_*^{j,t}$ can be seen as a particular linear combination of the 
output values $(y_d^{i,t})_{1\leq i\leq n_{ob}}$ belonging to the database. 
The full time-dependent evolution of the deceleration force profile is then given by $F_*=(f_*^{i,t})_{1\leq i \leq n_{test}, \; 0\leq t\leq T}$. 
We thus have built a purely statistical regression model from the database $D$ we have at our disposal: \\
$$\mathcal{R}_{GP}^{D}: \left\{
\begin{array}{ccc}
 \mathbb{R}^{n_{test}\times 6(T+1)} &\rightarrow &\mathbb{R}^{n_{test} \times (T+1)}\\
 X_* & \mapsto & F_*=(f_*^{i,t})_{1\leq i \leq n_{test}, \; 0\leq t\leq T}.\\
\end{array}
\right.
$$ 
From a training database $D$, and a set of new random onput vectors $X_* = (x_*^i)_{1\leq i \leq n_{test}} \in \mathbb{R}^{n_{test}\times 6(T+1)}$, this regression model 
enables to reconstruct the profile of the deceleration force of the plane during the landing $f^i_* = (f_*^{i,t})_{0\leq t \leq T}$ for the value of the input quantities $x^i_*$ ($1\leq i \leq n_{test}$).

\subsection{Fitting the hyperparameters: maximizing the marginal likelihood}\label{sec 2.3}

Let us denote by $\theta=(\sigma,\tau, l_k,\, 1\leq k\leq 6)\in\mathbb{R}^8_+$ a set of hyperparameters for the Bayesian gaussian process model introduced in 
Section~\ref{sec 2.1}, and let us denote by $K_{\textbf{X},\textbf{X}}(\theta)$ the random matrix defined by~\eqref{mu} using the kernel function $\kappa$ defined by~\eqref{ker}
 with this set of hyperparameters.\\
In this section, we present how we choose the value of these hyperparameters (which a priori depends on the time $t$ considered), 
$\theta_{opt}^t$, which we use in order to build the regression model for the reconstruction of the deceleration force profile of the plane, as described in 
Section~\ref{sec 2.2}.\\
The probability density functions of the random variables $\textbf{Y}^t|\textbf{F}^t,\textbf{X}$ and $\textbf{F}^t|\textbf{X}$ are functions which depend on the value of 
these hyperparameters and we denote then respectively by $p(\textbf{Y}^t|\textbf{F}^t,\textbf{X};\theta)$ and $p(\textbf{F}^t|\textbf{X};\theta)$. The probability density 
function of the variable $\textbf{Y}^t|\textbf{X}$ is called the \textit{marginal likelihood}, depends also on the value of the hyperparameters $\theta$ and can 
be expressed as a function of the prior and likelihood distributions
$$p\big ( \textbf{Y}^t|\textbf{X};\theta \big )=\int p\big ( \textbf{Y}^t|\textbf{F}^t=F^t\textbf{X};\theta \big )p(\textbf{F}^t=F^t|\textbf{X};\theta) dF^t.$$
       
Using the gaussian process model described in the preceding section, we can derive an explicit expression of the  \textit{{\rm log}  marginal likelihood} 
${\rm log} \;p(\textbf{Y}^t|\textbf{X};\theta)$ as follows (see~\cite{ed}):
$${\rm log}\; p(\textbf{Y}^t|\textbf{X};\theta)=-\frac{1}{2}\textbf{Y}^t\big ( K_{\textbf{X},\textbf{X}}(\theta)+\sigma^2 I_{n_{ob}} \big )^{-1}\textbf{Y}^t--\frac{1}{2}{\rm log}\big ({\rm det}(K_{\textbf{X},\textbf{X}}(\theta)+\sigma^2 I_{n_{ob}})  \big )-\frac{n_{ob}}{2}{\rm log}(2\pi)$$   
       
A classical approach to set the values of the hyperparameters for a given time $0 \leq t\leq T$ in an optimal way is to maximize the marginal likelihood of the 
database we have at our disposal, in other words, $\theta^t_{opt}$ is chosen to be solution of
$$\theta_{opt}^t\in \mathop{\mbox{argmax}}_{\theta\in \mathbb{R}^8_{+}}\mathcal{L}^t(\theta),$$      
where        
\begin{align*}
\mathcal{L}^t(\theta) & :=  {\rm log}\;p\big ( \textbf{Y}^t=Y^t_d|\textbf{X}=X_d; \theta \big ) \\
  & =  -\frac{1}{2}Y_d^t\big (  K_{X_d,X_d}(\theta)+\sigma^2 I_{n_{ob}}  \big )^{-1}Y_d^{t} -\frac{1}{2}{\rm log}\big ( {\rm det}( K_{X_d,X_d}(\theta) +\sigma^2 I_{n_{ob}})  \big )-\frac{n_{ob}}{2}{\rm log}(2\pi).
\end{align*}  
Thus, this set $\theta_{opt}^t$ of hyperparameters is chosen to be the one which makes the database we have "as likely as possible".\\
In principle, the values of the hyperparameters $\theta_{opt}^t$ should be computed for each time $0 \leq t\leq T$, which would lead to the resolution of
 $(T+1)=101$ optimization problems depending on $8$ parameters each. This ideal approach is too costly from a computational point of view, so we adopted a 
simplified approach which requires however to make additional assumptions on the law of the process $(\textbf{Y}^t)_{0 \leq t\leq T}$.\\
For $N\in \mathbb{N}^*$, let us introduce $T_1,T_2,...,T_N\in\{0,...,T\}$ such that
$$\widetilde{T}_1:=0<T_1<\widetilde{T}_2:=T_1+1<T_2<\widetilde{T}_3:=T_2+1<...<\widetilde{T}_{N-1}:=T_N+1<T_N=T.$$
 Instead of computing different sets of  hyperparameters $\theta_{opt}^t$ for all times $0 \leq t\leq T$, we only compute for all $0 \leq m\leq N$, one set of hyperparameters
 $\theta_{opt}^m$ which will be the same for all times $t$ belonging to the time subinterval $[\widetilde{T}_m,T_m]$. To compute the optimal value $\theta_{opt}^m$ of this set of  hyperparameters, we make an additional assumption on the law of $(\textbf{Y}^t)_{0 \leq t\leq T}$: we assume that for all $1 \leq m\leq N$, the random vectors $\textbf{Y}^t|\textbf{X}$ are independent from one another for all $t\in [\widetilde{T}_m,T_m]$. This implies that we assume that there is no correlation between the values of the observed output quantities at two different times belonging to the same time subinterval $[\widetilde{T}_m,T_m]$, which is not true in general of course. However, this very crude assumption enables us to significantly simplify the calculations of the hyperparameters while the produced regression model gives very reasonable results as will be seen in Section \ref{sec 3}.\\
 For all $1 \leq m\leq N$, the optimal value of the hyperparameters $\theta_{opt}^m$ is then chosen as the solution of the following optimization problem
 \begin{equation}\label{thet}
 \theta_{opt}^m\in \mathop{\mbox{ argmax}}_{\theta\in \mathbb{R}^8_{+}}\mathcal{L}^m(\theta)
\end{equation}     
where 
\begin{align*}
\mathcal{L}^m(\theta) & :=  \displaystyle \sum_{t=\widetilde{T}_m}^{T_m}\mathcal{L}^t(\theta) \\
  & = \displaystyle \sum_{t=\widetilde{T}_m}^{T_m}  {\rm log}\;p\big ( \textbf{Y}^t=Y^t_d|\textbf{X}=X_d; \theta \big ) \\
    & = \displaystyle \sum_{t=\widetilde{T}_m}^{T_m} -\frac{1}{2}Y_d^t\big ( K_{X_d,X_d}(\theta)+\sigma^2 I_{n_{ob}} \big )^{-1}Y_d^t  -\frac{(T_m-\widetilde{T}_m+1)}{2}{\rm log}\big ( {\rm det}( K_{X_d,X_d} ( \theta) +\sigma^2 I_{n_{ob}})  \big )-\frac{(T_m-\widetilde{T}_m+1)}{2}{\rm log}(2\pi).
\end{align*}         
    In the numerical results presented in Section~\ref{sec 3}, we illustrate two different ways to choose the times $T_1,...,T_N$:
    \begin{itemize}
    \item a first choice consists in taking $N=1$, and thus $\widetilde{T}_1=0$ and $T_1=T$; in this case, we only compute one set of hyperparameters 
$\theta_{opt}^1=(\sigma_{opt}^1,\tau^1_{opt},l_{k,opt}^1,\,1\leq k\leq 6)$ which are valid for the reconstruction of the deceleration profile of the plane at all times
 $0 \leq t\leq T$;
    \item a second choice, consists in taking $N=10$ and for all $1 \leq m\leq N$, $T_m=10 m$; the true interval $[0,T]$ is partitioned into $10$ time subintervals 
and we compute $10$ sets of optimal hyperparameters for each of these subintervals. 
    \end{itemize}       
\noindent In Section~\ref{sec 3}, we compare the results obtained with the first and second strategy. The optimization problems~\eqref{thet} are solved in practice 
using a standard gradient algorithm.

 \section{Numerical tests}\label{sec 3}

  \subsection{Presentation of other statistical models}\label{sec 3.1}

As mentioned before, we compared the approach we detailed in Section~\ref{sec 2}. To this aim, we consider the following different strategies:
\begin{itemize}
 \item linear regression (LR); 
\item generalized additive model (GAM);
\item multivariate adaptative regression splines (MARS); 
\item random forests (RF).
\end{itemize}
We denote respectively by $\mathcal{R}_{LR}^{D}$, $\mathcal{R}_{GAM}^{D}$, $\mathcal{R}_{MARS}^D$ and $\mathcal{R}_{RF}^D$ 
the obtained regression models with the training database $D$, which are all applications from $\mathbb{R}^{n_{test}\times 6(T+1)}$ to $\mathbb{R}^{n_{test}\times (T+1)}$. 
Let us present the general idea of each of those models except for the well-known linear regression, using the notation of the preceding section. For the sake of brievity, we 
do not give all implementation details here. \\

The generalized additive model (see~\cite{gen}) is an extension of the generalized linear regression approach to non-linear relationships proposed by Hastie and Tibshirani. 
This model is constructed as a sum of smooth functions of each of the covariates. The interest of such a method is that each smooth function is able to reproduce any shapes. 
The smooth functions were estimated by cubic regression splines. In other words, for all $1\leq i \leq n_{test}$, if 
$x^i_*=(x_{*,k}^{i,t})_{1\leq k \leq 6, \; 0 \leq t \leq T}\in \mathbb{R}^{6(T+1)}$, then the associated deceleration profile $(f_*^{i,t})_{0\leq t \leq T}$ is reconstructed as follows
\begin{equation}\label{gen}
f_*^{i,t} = \beta_0^D + \displaystyle \sum_{k=1}^6h^D_k\left(x_{*,k}^{i,t}\right),
\end{equation}
where $h^D_k$ is a smooth function for all $1\leq k \leq 6$ and $\beta_0^D\in \mathbb{R}$.\\

MARS (see~\cite{mul}) is an automatic procedure for modeling the output using the most significant non-linear relationships and interactions between covariates.
 The model is a sum of basis functions which are either a single hinge function or either a product of one or more hinge functions~\eqref{mars}. A hinge function is a 
piecewise function with two pieces on both sides of a knot. One piece is set at zero and the other piece corresponds to a linear function. 
For all $x^i_*=(x_{*,k}^{i,t})_{1\leq k \leq 6 ,\;0 \leq t \leq T}\in \mathbb{R}^{6(T+1)}$, then $(f_*^{i,t})_{0\leq t \leq T}$ is reconstructed as follows
\begin{equation}\label{mars}
f_*^{i,t}=\beta_0^D+\displaystyle \sum_{m=1}^M \beta_m^D h_m^D(x_{*,1}^{i,t},x_{*,2}^{i,t},...,x_{*,6}^{i,t}),
\end{equation}
where $(\beta_m^D)_{1\leq m\leq M }\in \mathbb{R}^M$ are the intercept and the slope parameters and $(h^D_m)_{1\leq m\leq M }$ being $m$ smooth real-valued functions 
defined on $\mathbb{R}^6$.
MARS automatically selects the most significant basis functions by applying a procedure with two steps: a forward pass and a pruning pass. The forward pass delivers a model
 with too many basis functions that overfit the data while the pruning pass removes the least significant basis functions to obtain the most accurate sub-model. 
The forward pass selects iteratively the best reflected pair of hinge function among all possible functions. The set of possible functions is built taking all observed 
covariates values as a knot. Then, the pruning pass removes the least significant basis function one by one until it finds the most accurate subset of basis functions. \\

Finally, random forests (see~\cite{for}) is a machine learning algorithm. Such a model is composed by an ensemble of decision tree models built on 
random learning datasets. Briefly, a tree model is a recursive partitioning of the observations according to their similarities in covariates and output. Tree models
 are built using the classification and regression tree (CART) algorithm. The initial dataset is split into two clusters according to a threshold value for one 
covariate: one cluster have higher value or the other cluster lower value. The algorithm evaluates all possible thresholds and selects the one which minimizes
 the total sum of squared errors. The partitioning is repeated for each cluster until there are less than $5$ observations per cluster. The random forest algorithm 
creates $500$ tree models from $500$ random samples of the original dataset. The output estimated by random forest is an average of the individual 
estimation of all tree models.

  \subsection{Model validation}\label{sec 3.2}

To assess the validity of a regression model $\mathcal{R}$ such as the ones presented in Section~\ref{sec 2} or in Section~\ref{sec 3.1}, 
we perform \textit{cross-validation} tests. The principle is the following: the full database $D$ we have at our disposal is composed of $n_{ob}=1505$ 
different landings, so that $D=\{(y_d^i,x_d^i),\;i\in \mathcal{I}_{tot}:=\{1,...,n_{ob}\}\},$ using the notation of Section~\ref{sec 1.4}. 
A regression model $\mathcal{R}^D$ (which can be built for instance using the Bayesian approach with gaussian processes we detailed in Section~\ref{sec 2}) 
depends of course of the information contained in $D$. 
Cross-validation tests consist in training a regression model from a smaller database than the one we have access to, and, 
for all the recordings which were drawn from the training database, to compare the profile of the output quantities reconstructed from the statistical 
regression model and the measured output profile. 
More precisely, let $\mathcal{I}_1,\mathcal{I}_2,...,\mathcal{I}_M \subset \mathcal{I}_{tot}$ be $M$ disjoint sets of recording indices. 
For all $1\leq p \leq M$, we denote by $\mathcal{R}^{D_p}$ the regression model built from the database $D_p$ using one of the strategies presented in Sections~\ref{sec 2} 
or~\ref{sec 3.2}, where
$$D_p:=\{(y_d^i,x_d^i),\;i\in\mathcal{I}_{tot} \setminus \mathcal{I}_p \}.$$
Let us assume for the sake of simplicity that for all $1\leq p \leq M$, $\mbox{Card}\mathcal{I}_p = n_{test}$. 
Assessing the validation of a regression model $\mathcal{R}$ amounts to comparing the error between the measured output profiles 
$(y_d^i)_{i\in \mathcal{I}_p} \in \mathbb{R}^{n_{test}\times (T+1)}$ and the reconstructed output 
profile $F_*=(f_*^i)_{i\in \mathcal{I}_p}=\mathcal{R}^{D_p}(X_*)$ taking $X_*:= (x_d^i)_{i\in \mathcal{I}_p}\in \mathbb{R}^{n_{test}\times (6(T+1))}$ as a test input set.

  \subsection{Numerical results}\label{sec 3.3}
  
We performed cross-validation tests with these five different models using $M=10$ training data sets such that 
for all $1\leq p\leq M$, ${\rm Card}\;\mathcal{I}_p=n_{test}=150$. For the Gaussian process model, we tried two different strategies for 
the splitting of the time interval for the fitting of hyperparameters as mentioned in Section~\ref{sec 2.3}. In Section~\ref{sec 3.3.1}, 
we compared the different regression strategies using only the gaussian process model reconstructed with $N=1$ time subinterval. 
In Section~\ref{sec 3.3.2}, we compare the results obtained for $\mathcal{R}_{GP}$ with $N=1$ or $N=10$ subintervals.

The curve legen is the following:
\begin{itemize}
\item the black dots curve refers to the trye measured data; 
\item the black line curve refers to the Gaussian process model;  
\item the black dashed line curve refers to the linear regression,
\item the black dotted line curve refers to the generalized additive model;
\item the grey line curve refers to the Random Forest strategies;
\item the grey dashed line curve refers to the Multivariate Adaptative Regression Splines (MARS).
\end{itemize}

  \subsubsection{Comparison of the different regression models}\label{sec 3.3.1}
  
We apply the five regression strategies on this problem and we draw the curve of each estimated deceleration profile on two different landings, 
the $23^{th} $and the $45^{th}$ landing for example (see Figure 1), but only on the first $40$ seconds. This is the time period when the influence of the braking system 
is the most important during the landing.

     \begin{figure}[htp]
  \centering

  \begin{tabular}{cc}

    \includegraphics[width=75mm]{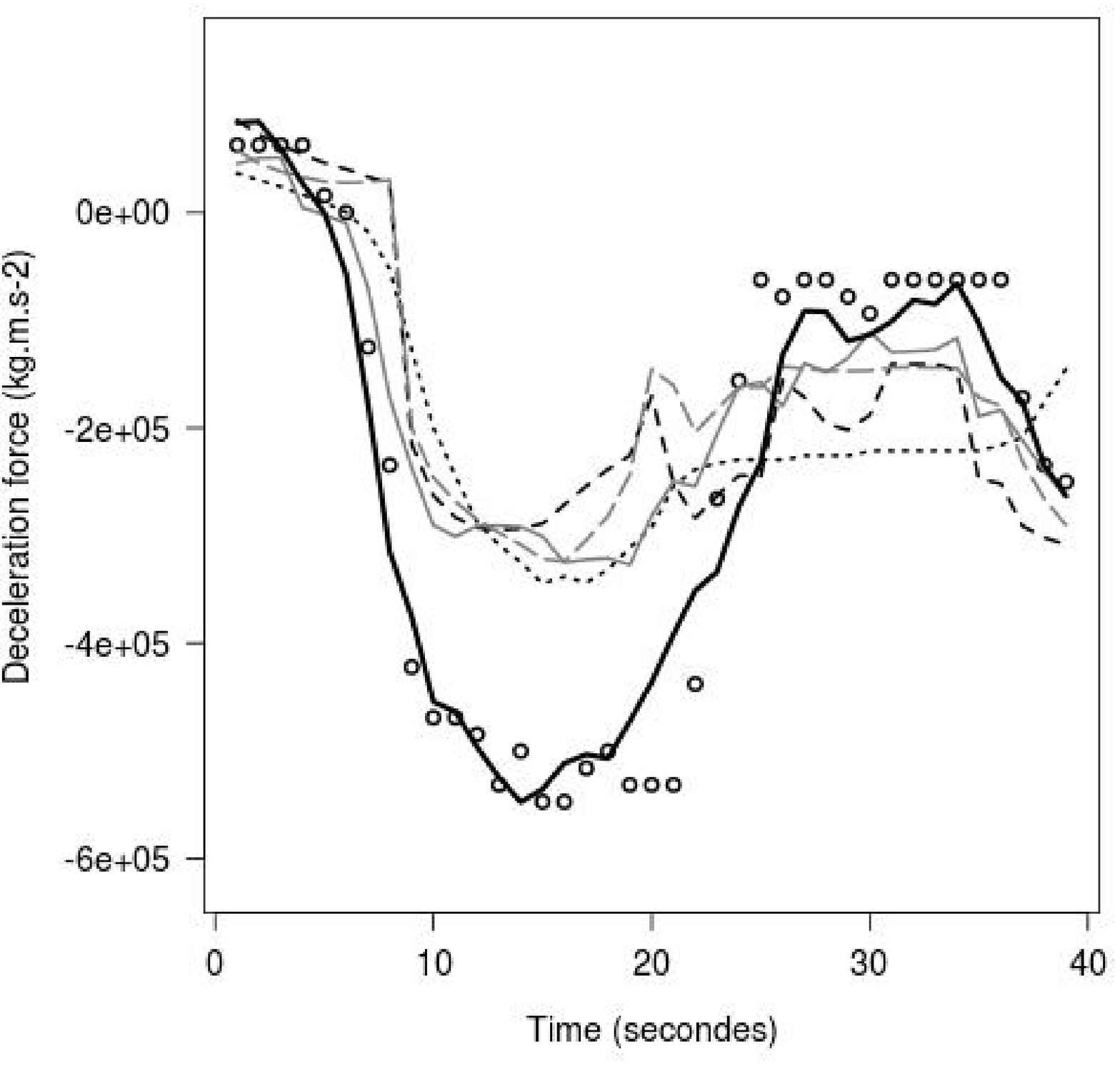}&\label{2lan}

    \includegraphics[width=75mm]{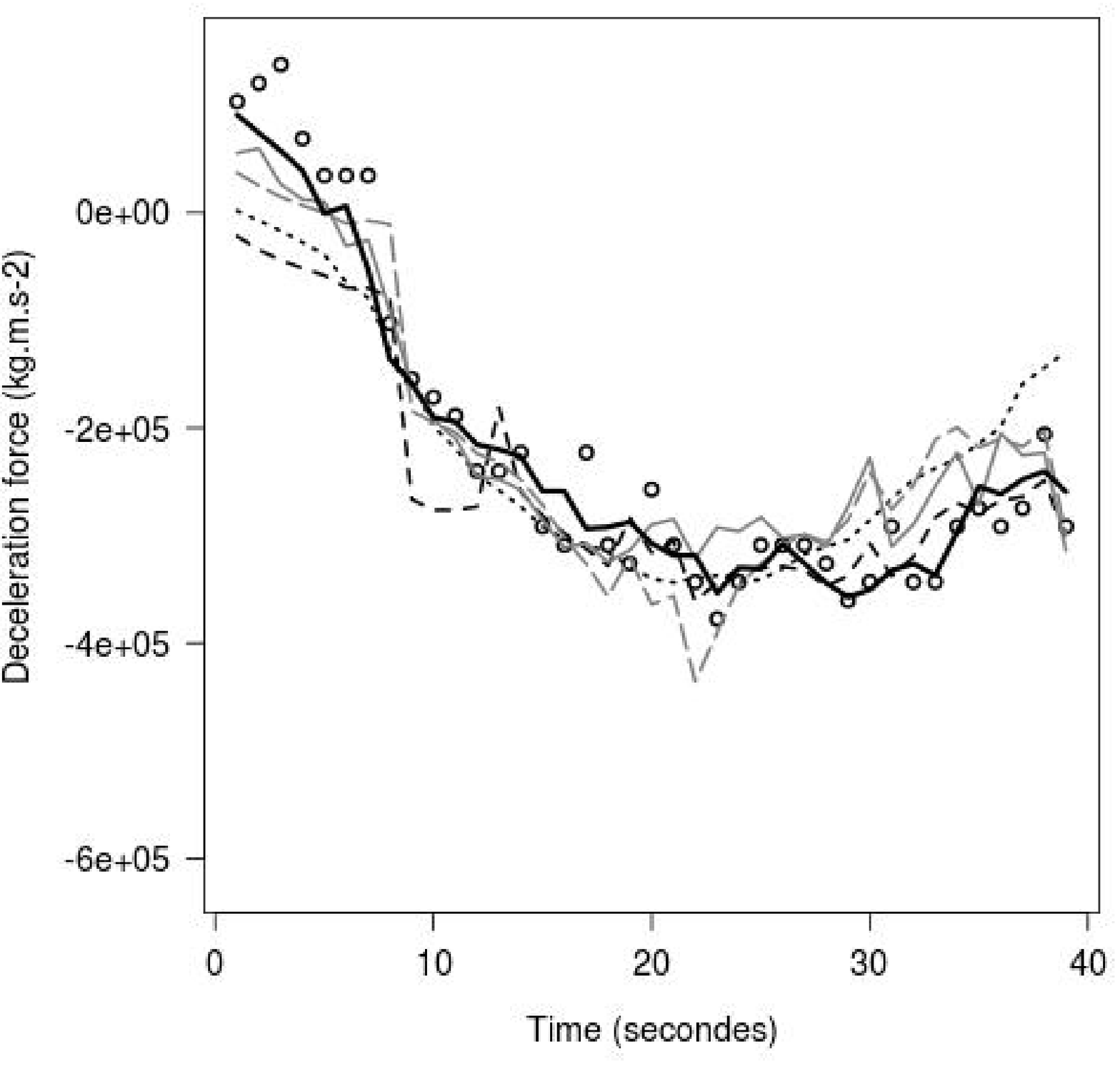}\\

  \end{tabular}
\tiny\caption{\small{Data curves predicted for the $5$ models and the observed curve for landing line $23$(left) and landing line $45$(right) of validation set.}}

\end{figure}

 We can clearly see on these two examples that the Gaussian Process approach is the one that fits best the measured deceleration profiles. 
This was observed for the majority of the landing profiles considered.

In Figure~2, we plot the median absolute percentage error (MAPE) for all models, which is defined by the following formula (using the notation of Section~\ref{sec 3.2}):\\ 
    \begin{equation}
    MAPE=\displaystyle\frac{1}{M}\sum_{i=1}^{M}\frac{1}{\# \mathcal{I}_p }\sum_{i\in \mathcal{I}_p} \frac{\|y_d^i-f_*^i\|_{\mathbb{R}^{(T+1)}}}{\|y_d^i\|_{\mathbb{R}^{(T+1)}}} 
        \end{equation}
From this criterion, it can be seen that the gaussian process approach we propose significantly improves the predictive quality of the regression models 
previously used by Safety Line.
     \begin{figure}[htp]
  \centering
\includegraphics[width=75mm]{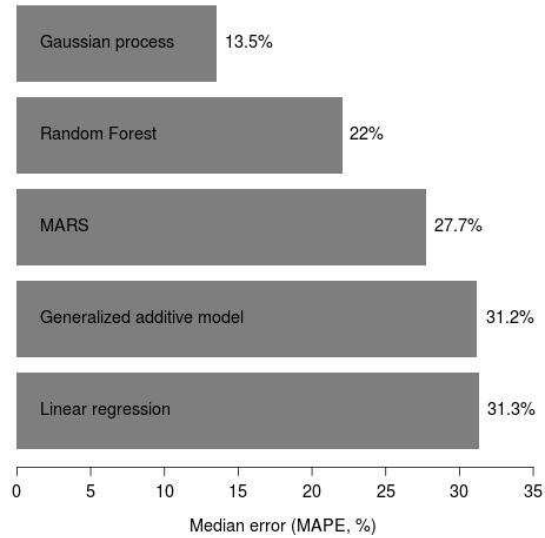}\label{eror}
\tiny\caption{\small{The median absolute percentage error for the five models.}}
\end{figure}
\newpage

The histogram of the number of errors plotted in Figure~3 provides another criterion to compare the different regression models. 
More precisely, for all $s\in \{0,...,99\}$, the histogram plots the cardinal of the set 
$$ \displaystyle \bigcup_{p\in \{1,...,M\}}\left\{ i\in \mathcal{I}_p,\; s \% \leq \frac{\|y_d^i-f_*^i\|_{\mathbb{R}^{(T+1)}}}{\|y_d^i\|_{\mathbb{R}^{(T+1)}}} \leq (s+1)\%  \right \}.$$
     \begin{figure}[htp]
  \centering
\includegraphics[width=75mm]{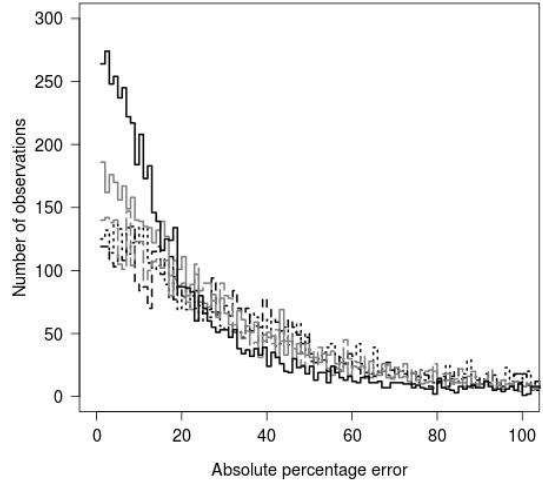}\label{hist}
\tiny\caption{\small{The histogram of errors of each model in the first $40$ seconds of landing.}}
\end{figure}

Of course, a high-quality model will produce a large number of errors for small values of the error threshold $s$ and a small number of errors for large values of $s$. This is 
indeed the case for the Gaussian process model, and we can see that it also performs significantly better than the other regression models from this point of view.  \\

% A possible explanation of the significant improvement of the results using our Gaussian process based approach compared with the other methods we presented in this proceeding may be the following.\\
% Using process based on Gaussian Process, a new output signal is defined as a linear combination of other signals that are already present in the database. However, other methods reconstruct the output signal as a linear combination of input signals, which are not of the same nature as the output signal, and may present different behavior of time correlation effects. This may account for the fact that the time dependencies of the output signal are qualitatively well reproduced in our approach, even if we use such a naive way to incorporate time correlation effects in our statistical model. From this point of view, Gaussian process statistical methods seem share some common features with the Reduced Basis method.  

  \subsubsection{Influence of time subintervals splitting with the gaussian process approach}\label{sec 3.3.2}
  
Let us now compare the results we obtained with the Gaussian Process model presented in Section~\ref{sec 2}, 
where we used only $N=1$ time interval, or $N=10$ different time subintervals.\\
In this figure, we compare the error associated to these two different strategies, for each subintervals of $10$ seconds. 
In other words, for each $1\leq n \leq N $, we compute the MAPE error 
$$    MAPE_n=\displaystyle\frac{1}{M}\sum_{i=1}^{M}\frac{1}{\# \mathcal{I}_p }\sum_{i\in \mathcal{I}_p} \frac{\left (\displaystyle\sum_{t=\widetilde{T}_n}^{T_n}
\left (y_d^{i,t}-f_*^{i,t}\right )^2\right )^{\frac{1}{2}}}{\left (\displaystyle\sum_{t=\widetilde{T}_n}^{T_n}\left (y_d^{i,t}\right )^2\right )^{\frac{1}{2}}}$$
for the two strategies, where $\widetilde{T}_n = 10(n-1)$ and $T_n = 10n$.\\
As expected, using different time subintervals improves the results, especially in the last part of landing.

     \begin{figure}[htp]
  \centering
\includegraphics[width=75mm]{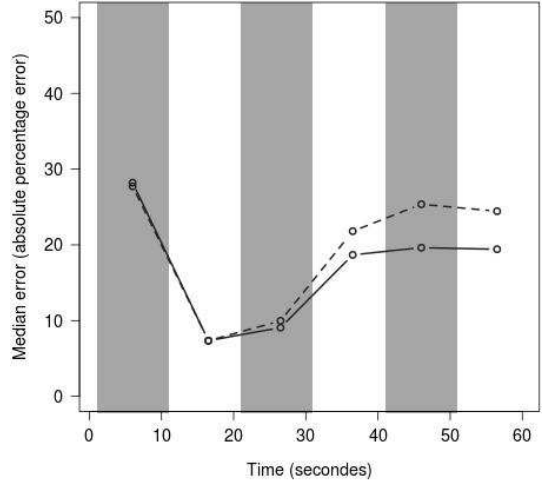}\label{tran}
\tiny\caption{\small{The median error observed for each section of $10$ seconds for the model with $N=1$ (dashed curve) and $N=10$ (full curve).}}
\end{figure}
\newpage

\section{Interpretation of the results and discussion about time correlations}\label{disc}

In this section, we wish to comment the way time correlations are taken into account in our model described in Section~\ref{sec 2} through the very simplistic choice of the 
kernel $\kappa$ we used here (\ref{ker}). Let us first note that it is quite usual in 
time-dependent regression models that the output quantity to be reconstructed at time $t$, $\textbf{y}^t$, usually depends only on the value at time $t$ 
of an input quantity $\textbf{x}^t$ (see~\cite{mod} for instance). However, this is clearly not the case here since the deceleration of the place at 
an instant $t$ depends a priori on all the set of past input values $(\textbf{x}^{t'})_{0\leq t' \leq t}$. The form of the kernel function $\kappa$ 
enables to take into account in some way the fact that the value of the ouput quantity $\textbf{y}^t$ at time $t$ depends on the whole trajectory 
$(\textbf{x}^{t'})_{0\leq t' \leq t}$. The dependence of the value of $\textbf{y}^t$ on the different values $(\textbf{x}^{t'})_{0\leq t'\leq T}$ 
is somehow aggregated through the use of the Frobenius norm $\|\cdot\|_F$, which is of course a very naive approach. 
One shortcoming of this model in particular is that the value 
$\textbf{y}^t$ then depends on the \itshape future \normalfont values of the input quantities, which is of course unrealistic from a physical point of view.

\medskip

This very simplistic approach is sufficient to yield very satisfactory numerical results though and seems to capture somehow some features of the 
time correlations between input and output data. 
A possible explanation of the significant improvement of the results using our Gaussian process based approach compared with the other methods 
we presented in this proceeding (which do not take time correlation into account as well) may be the following.\\

In our approach, a new output signal is reconstructed as a linear combination of other signals that are already present
in the database. 
However, other methods reconstruct the output signal as a linear combination of input signals, which are not of the same nature as 
the output signal, and may present different behaviour of time correlation effects. 
This particularity of the gaussian process based approach may account for the fact that the time dependencies of the output signal 
are qualitatively well reproduced in our case, 
even if we use such a naive way to incorporate time correlation effects in our statistical model.

\medskip

In the rest of the section, let us comment on different strategies which could have been used to incorporate time correlation effects using gaussian processes approach in our regression model. 
A first strategy could have been to consider the time $\textbf{t}$ as a random variable, in the same way as the input quantity 
$\textbf{x}$. This would require to modify the mean function $\mu$ and kernel function $\kappa$ so that 
they do not only depend on values of the input quantities $\textbf{x}$ but also on time. 
More precisely, following ideas of~\cite{spa}, the output quantity $\textbf{y}^{\textbf{t}}$ could be modeled by
$$
\textbf{y}^{\textbf{t}} = \mathcal{F}(\textbf{x}, \textbf{t}) + \omega(\textbf{x}, \textbf{t}), 
$$
where $\mathcal{F}: \mathbb{R}^{6(T+1) \times (T+1)} \to \mathbb{R}$ would be a random function to be determined by the regression model and $\omega(\textbf{x}, \textbf{t})$ 
a white noise random process for instance. Using similar notation as 
those used above, the random value $\textbf{f}^{\textbf{t}} = \mathcal{F}(\textbf{x}, \textbf{t})$ could be modeled as a gaussian process characterized by a mean 
$\mu: \mathbb{R}^{6(T+1) \times (T+1)}\to \mathbb{R}$ and a covariance kernel $\kappa: \mathbb{R}^{6(T+1) \times (T+1)}\to \mathbb{R}$ so that the law of 
$\textbf{f}^t|\textbf{x}, \textbf{t}$ would be given by
$$
\textbf{f}^{\textbf{t}} | \textbf{x}, \textbf{t} \sim \mathcal{GP}(\mu(\textbf{x}, \textbf{t}), \kappa(\textbf{x}, \textbf{t}; \textbf{x'}, \textbf{t'})).
$$
Such an approach would enable to take into account correlations between values of the output and input quantities at different times in a natural way. 
The difficulty we encounter with such an approach is that the standard reconstruction procedure derived from this gaussian process approach requires the inversion 
of a matrix of size $(n_{ob}\times (T+1)) \times (n_{ob} \times (T+1))$ where $n_{ob} \times T+1 \approx 150 000$ in our case. 
The large size of this matrix makes its inversion very difficult from a practical point of view. 
However, in~\cite{spa}, the authors proposed to convert such a \itshape spatio-temporal \normalfont Gaussian process 
into an infinite-dimensional Kalmann filtering. The interest 
of such an approach is that the complexity of such an approach is linear (instead of cubic) in the number of time steps, thus avoiding the numerical difficulties 
mentioned above. We did not test this strategy in our case though.

\medskip

However, a second (more tractable) way to improve the choice of this kernel function could rely in the modification 
of the kernel function $\kappa$ in order to take into account time correlatiosn as follows. 
Indeed, using the squared exponential kernel function~\eqref{ker}, in order to reconstruct the value of the deceleration of the plane at a time $t$, all the values of the input quantities at
 all times have the same importance, which is of course unrealistic. One would reasonably expect that that only the values of the input quantities at times anterior to $t$ would affect the value of the output quantity at a time $t$. 
 
To take this into account, one could think of using at each time $t$ a modified kernel function $\kappa^t$ defined as follows:\\
for all $x^r =(x^{r,t}_k)_{0\leq t\leq T,\;1\leq k\leq 6}\in\mathbb{R}^{6(T+1)},$ $x^s =(x^{s,t}_k)_{0\leq t\leq T,\;1\leq k\leq 6}\in\mathbb{R}^{6(T+1)}$,
\begin{equation}\label{ker2}
\kappa^t(x^r,x^s):=\tau^2 {\rm exp}\left [  \displaystyle \sum_{k=1}^6\frac{1}{2l_k} \|x^{r,.}_{k}-x^{s,.}_{k}\|^t_{\mathbb{R}^{(T+1)}} \right]
\end{equation}
where the standard Frobenius norm $\|\cdot\|_{\mathbb{R}^{(T+1)}}$ would be replaced by a modified semi-norm 
$\|\cdot\|_{\mathbb{R}^{(T+1)}}^t$, which would depend on $t$ and could be written as follows:  for all $x=(x^{t'})_{0\leq t'\leq T} \in\mathbb{R}^{T+1}$,
$$\|x\|^t_{\mathbb{R}^{(T+1)}} :=\displaystyle \sum_{t'=0}^T w^t(t')|x^{t'}|^2,$$
 with a weight function $w^t:\{0,...,T\} \rightarrow \mathbb{R}_{+}$, satisfying $w^t(t')=0$ for all $t'\in \{t+1,...,T\}$ (the standard Frobenius norm used 
in~\eqref{ker} corresponds to $w^t\equiv 1$ for all $t\in \{0,...,T\}$). 
It would be interesting to test if these modifications could improve the quality of our regression model. This strategy would lead to an additional computational cost though: 
it woud require the storage of $T+1$ matrices of size $n_{ob}\times n_{ob}$ corresponding to all the matrices
$$
K^t_{\textbf{X},\textbf{X}} =(\kappa^t(\textbf{x}^i,\textbf{x}^j) )_{1\leq i,j \leq n_{ob}}, \quad \forall 0\leq t \leq T.
$$
Even if such a procedure could be more easly implementable than the first approach we described, we did not test this strategy here. It would be 
interesting though to check if one of these two possible strategies could help improving the numerical results presented in Section~\ref{sec 3.3}.

\section{Acknowledgments}

The authors thank Tony Leli\`evre and Nicolas Chopin for very helpful discussions. Labex AMIES is granted for financial support. This work was done in a summer school "CEMRACS 2013". We would like to thank the organizers of this school. Finally, Houssam Alrachid would like to thank "ENPC" and "CNRS Libanais" for supporting his PHD thesis.

%%-----------------------------
%%      your bibliography
%%-----------------------------

\end{document}